\documentclass{emulateapj}

\usepackage{amsmath}

\begin{document}

\title{Computing High Accuracy Power Spectra with Pico}

\author{William A.~Fendt\altaffilmark{1} and
        Benjamin D.~Wandelt\altaffilmark{1,2,3}}

\altaffiltext{1}{Department of Physics, UIUC, 1110 W Green Street,
             Urbana, IL 61801; fendt@uiuc.edu}
\altaffiltext{2}{Department of Astronomy, UIUC, 1002 W Green
             Street, Urbana, IL 61801; bwandelt@uiuc.edu}
\altaffiltext{3}{Center for Advanced Studies, UIUC, 912 W Illinois
             Street, Urbana, IL 61801}


\begin{abstract}

This paper presents the second release of Pico (Parameters for the Impatient 
COsmologist). 
Pico is a general purpose machine learning code which we have applied
to computing the CMB power spectra and the WMAP likelihood.
For this release, 
we have made improvements to the algorithm as well as the data 
sets used to train Pico,
leading to a significant improvement in accuracy.
For the $9$ parameter nonflat case presented here Pico can on average compute the 
TT, TE and EE spectra to better than $1\%$ of cosmic standard deviation
for nearly all $\ell$ values over a large region of parameter space.
Performing a cosmological parameter analysis of current CMB and large scale
structure data, we show that these power spectra give very accurate $1$ and $2$ 
dimensional parameter posteriors.
We have extended Pico to allow computation of the tensor power 
spectrum and the matter transfer function.
Pico runs about $1500$ times faster than CAMB at the default accuracy and
about $250,000$ times faster at high accuracy.
Training Pico can be done using massively parallel computing resources, 
including distributed computing projects such as Cosmology@Home.
On the homepage for Pico, located at
\verb+http://cosmos.astro.uiuc.edu/pico+,
we provide new sets of regression coefficients and make the training
code available for public use.
\end{abstract}



\section{Introduction}\label{intro}
Given the quantity of data available from current experiments such as WMAP and SDSS as well 
as the prospects for the next generation Planck and DES experiments on the horizon,
there is growing need for cosmologists to develop
tools that can accurately interpret the flood of data these experiments will gather.
A key component in all such analysis is the exploration of the posterior
density of the cosmological parameters given the available data.  This allows 
us to constrain and test theoretical models of the Universe.
A major computational hurdle in this procedure is the ability to quickly and accurately
compute the power spectrum of CMB fluctuations and the matter transfer function.  
This is accomplished using codes such as CMBFast \citep{Seljak:1996is} and 
CAMB \citep{Lewis:1999bs} 
that evolve the Boltzmann equation for the various constituents of the Universe.  
Using the default accuracy settings in CAMB the calculation of a single power spectrum
takes on the order of a minute. At higher settings, as may be required by upcoming
CMB experiments, the computational time can jump to several hours 
on a modern desktop. Furthermore, computation of constraints based on the data 
requires evaluating the power spectrum at $\mathcal{O}\left(10^5 - 10^6\right)$
models.
Decreasing the time required to calculate the CMB power spectrum, while maintaining 
sub-cosmic variance accuracy, will play an important part of
turning raw data into quantitative information about the history and structure of the 
Universe.

Previous codes aimed at speeding up power spectrum computations
such as DASH \citep{Kaplinghat:2002mh}  and CMBwarp 
\citep{Jimenez:2004ct} have attempted to reproduce the
computation of CMBFast or CAMB. 
More recently, motivated to develop a code that was both faster than DASH and
more accurate than CMBwarp, we released 
a machine learning code called Pico \citep{Fendt:2006uh}. It uses a 
training set of power spectra from CAMB to fit several multivariate 
polynomials as a function of the input parameters.  Along with accurately 
fitting the power spectra, we also found that
Pico was able to directly fit the WMAP likelihood 
\citep{Spergel:2006hy,Hinshaw:2006ia,Page:2006hz}.
This was done previously by CMBFit \citep{Sandvik:2003ii}
which used a similar idea of fitting the
likelihood with a polynomial in the cosmological parameters.
By replacing CAMB and the WMAP likelihood code with Pico we demonstrated that it 
can quickly explore the parameter space and give nearly identical posteriors.
Since the first release of Pico, Auld \textit{et al.} have applied a neural 
network code called CosmoNet to flat \citep{Auld:2006pm} and nonflat 
\citep{Auld:2007qz} models. 
Also Habib \textit{et al.} have demonstrated that a Gaussian process 
model introduced in \citep{Heitmann:2006hr} can 
predict the CMB power spectrum for flat models
based on a very small training set \citep{Habib:2007ca}.
Here we discuss some improvements to increase the accuracy of Pico 
and extend its to application to more general cosmological models.
Further, we describe a training method that avoids serial runs of CAMB
and is designed to leverage access to massively parallel and even 
distributed computing resources.  For example, we demonstrate that
Pico can be trained using the thousands of geographically distinct hosts
that contribute to Cosmology@Home.\footnote{http://www.cosmologyathome.org}

This paper is organized as follows. Section \ref{sec:algorithm} summarizes 
some of the improvements to the Pico algorithm and how training sets are generated.
In section \ref{sec:results} we demonstrate the performance of Pico in 
computing the power spectrum, matter transfer function and WMAP likelihood
for a $9$ parameter nonflat model.  We show that Pico can be used to 
quickly explore the parameter posterior for this model while
accurately reproducing the $1$ and $2$ dimensional marginalized distributions.
Lastly we summarize and conclude in section \ref{sec:conclusion}.

\section{The Algorithm} \label{sec:algorithm}
\subsection{Overview}
Given a training set of Cosmological parameters $\mathbf{x}$ and CMB power spectra
$\mathbf{y}$, Pico models the function $\mathbf{y}=f\left(\mathbf{x}\right)$ in
$3$ steps. First it compresses the power spectra using a 
Karhunen-Lo\`{e}ve \citep{Karhunen:1946,Loeve:1955,Tegmark:1994ed} technique. 
For $\ell_{\mathrm{max}}=3000$, the temperature and polarization spectra 
due to scalar and tensor perturbations can be compressed to $\sim 180$ total numbers. 
Next the algorithm clusters the input
parameters into non-overlapping regions. This is done using a $k$-means 
algorithm \citep{MacQueen:1967,Kirby:2001} or by choosing hyperplanes
to manually partition the space. 
Lastly, Pico models the function by fitting a least-squares polynomial
to the compressed power spectra over each cluster.  
Details of the algorithm can be found in the appendix of \citet{Fendt:2006uh}.

\subsection{Improved Fitting}
The first change to the algorithm is that it now uses the LAPACK libraries
\footnote{http://www.netlib.org/lapack/} to perform matrix decomposition. 
LAPACK makes the algorithm more stable, allowing the use of higher order 
polynomials. This gives Pico better fits at low $\ell$ ($<200$) and high 
$\ell$ ($>1500$).  This also makes clustering
significantly less important but still useful for improving the low 
$\ell$ accuracy as well as improving the fit to the WMAP likelihood directly.
In particular, we have found that partitioning the training set along values
of constant $\Omega_{\mathrm{k}}$ improves the low $\ell$ accuracy 
in the temperature power spectrum and matter transfer function, 
while partitioning in $\tau$, the 
reionization depth, gives a large improvement in computing the 
polarization power spectrum at low $\ell$.

\subsection{Decreasing Numerical Noise}\label{subsec:noise}
\begin{figure}
\begin{center}
   \epsscale{1.15} \plotone{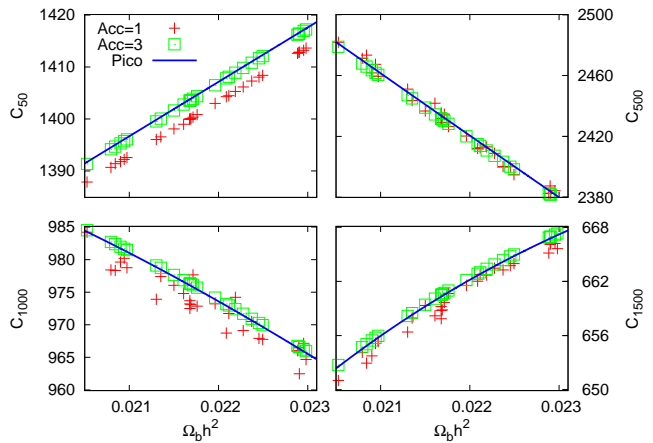}
   \caption{The plot shows the value of the temperature spectrum as a function of 
            $\Omega_{\mathrm{b}}h^2$ at various $\ell$-values for
            a nonflat cosmology. The red ($+$) points correspond to 
            to the default CAMB accuracies ($1$,$1$,$1$) and the
            green ($\Box$) points correspond to higher accuracy
            settings ($3$,$3$,$1$).
            While at low $\ell$ the power spectrum is smooth, the
            default accuracy becomes numerically noisy at higher $\ell$.
            This is one reason adding $\Omega_{\mathrm{k}}$ as a free
            parameter increases the difficulty in fitting the power
            spectrum. Also plotted, as a blue line, is the power
            spectrum computed by Pico trained on the $9$ parameter
            nonflat model discussed in section \ref{sec:results}.
            \label{fig:cl_noise}}
\end{center}
\end{figure}

\begin{figure}
\begin{center}
   \epsscale{1.15} \plotone{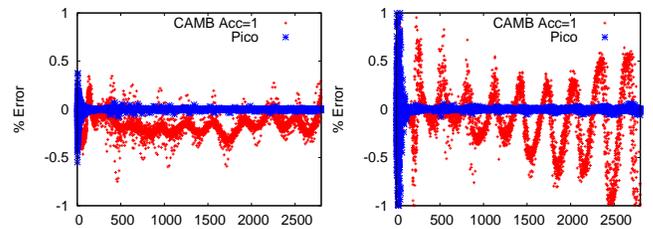} 
   \caption{The plots show the percent error between the TT (left) and
            EE (right) power spectrum computed by CAMB at the default 
            accuracy compared to those computed at high accuracy . 
            Also shown is the
            percent error between the power spectra computed by Pico
            and the high accuracy CAMB spectra.
            This test was done on $25$ models all located within $25$
            log-likelihoods of the WMAP peak.
            \label{fig:camb_acc}}
\end{center}
\end{figure}

For nonflat models, fitting algorithms 
are hindered by the fact that at the default accuracy the power spectra 
computed by CAMB are numerically noisy. 
This is demonstrated in Figure \ref{fig:cl_noise}, where
we have plotted the power spectrum 
at the default and high accuracy levels
for various $\ell$-values as a function of a 
single parameter ($\Omega_{\mathrm{b}}h^2$) for a nonflat model.
Since the power spectrum is not a numerically smooth function of the
cosmological parameters, Pico is limited in its ability to fit the
true spectra.
Also note that the higher accuracy power spectrum does not
always smooth over the lower accuracy case. 
To reduce this noise and limit the effect of interpolation errors
we have generated training sets with CAMB using the accuracy parameters 
set as: {\tt Accuracy=3}, {\tt lAccuracy=3} and {\tt lSamp=3}.
In Figure \ref{fig:camb_acc} we have plotted the error between
the default and high accuracy power spectra from CAMB for $25$ 
models around the peak of the WMAP likelihood. The left and right plots
show the percent error in the TT and EE spectra. Also shown is the error
between the spectra computed by Pico and the high accuracy CAMB runs.
Here Pico was trained on the $9$ parameter model discussed in
section \ref{sec:results}.
The figures demonstrate that when trained on high accuracy data Pico
computes the power spectrum around the peak of the likelihood to better 
precision than CAMB at its default accuracy settings.

\begin{figure}
\begin{center}
   \epsscale{1.15} \plotone{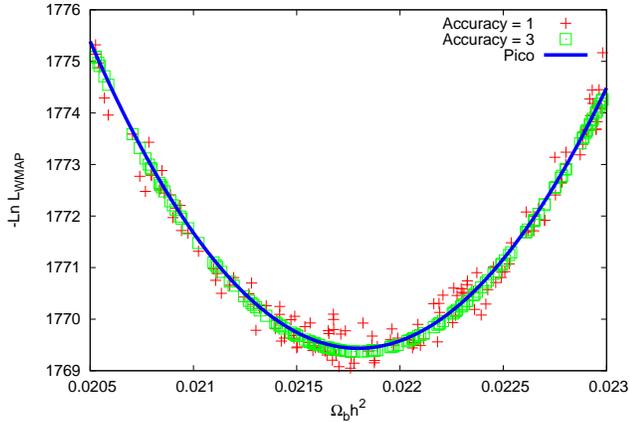}
   \caption{Value of $-\ln L_{\mathrm{WMAP}}$ as a function of
            $\Omega_{\mathrm{b}}h^2$ for a nonflat cosmology.
            Note that this is near the peak of the likelihood 
            in the full space. The red ($+$)
            points correspond to the default CAMB accuracies ($1$,$1$,$1$)
            and the green ($\Box$) points corresponding to using higher 
            accuracy settings ($3$,$3$,$1$). The blue line is the 
            value computed by Pico trained over the $9$ parameter nonflat
            models discussed in section \ref{sec:results}.
            Note that using the default accuracy in CAMB gives a numerically 
            noisy function, which can lead to variations of $1$ or more
            log-likelihoods, but Pico gives a smooth function through the high
            accuracy values.
            \label{fig:lnlike_noise}}
\end{center}
\end{figure}

Numerical noise in the power spectrum also
leads to noise in the WMAP likelihood as shown in Figure 
\ref{fig:lnlike_noise}.  
Again, this increases the difficulty in fitting the likelihood
with Pico.
Just as the power spectrum, this is remedied by running CAMB at
higher accuracy.
While the level of noise introduced in the likelihood may not be
of significant concern when analyzing current CMB data sets, it may 
represent an important hurdle to overcome for the next generation of 
experiments.
Also evident from Figure \ref{fig:lnlike_noise} is that Pico provides
a smooth approximation to the noisy likelihood. This is an important property
for algorithms that require differentiating the likelihood such as
Hamiltonian Monte Carlo \citep{Hajian:2006mt}.

\subsection{Generating the Training Set}\label{subsec:train}
As in the first release of Pico, we generate the training set of
power spectra and matter transfer function by sampling uniformly 
from a large box.
After training Pico to compute the power spectra, evaluation of the
WMAP likelihood, which requires a few seconds, becomes the new bottleneck
in parameter estimation. Another significant speed up can be obtained by 
using Pico to directly fit the likelihood function.  
However, for the $9$ parameter case we examine in the next section this 
is a difficult problem.
Training Pico over a box in parameter space 
includes regions that are many thousands of log-likelihoods from the peak
and gives a very sparse sampling of the high likelihood region.
In practice we are not interested in these areas of parameter space.
Instead we aim to compute the likelihood very accurately around the peak
of the distribution. This requires generating a training set for Pico that
includes only the high likelihood region.

A natural method of accomplishing this is to use the 
Metropolis Hastings algorithm to find points in the high likelihood region. 
This can be done efficiently by running CosmoMC \cite{Lewis:2002ah} and
using Pico to compute the power spectra. 
To ensure we cover a sufficient volume the chains are run 
using only the WMAP data and at a higher temperature, meaning the log-likelihood
is scaled by a constant factor allowing the chains to explore a larger volume.
This step is dominated
by the time it takes to run the WMAP likelihood code.
Lastly we run the samples through CAMB and the WMAP code
to get the true likelihood which will be used to retrain Pico. 
This step is also quick as it can easily be run in parallel.
It is useful to note that this procedure never requires running CAMB 
in serial making it an ideal application for distributed computing
projects such as Cosmology@Home.
The training set can be further refined by pruning out data at low
likelihood.
In the following Pico is trained to compute the likelihood on
points within $25$ log-likelihoods of the WMAP peak.

\subsection{Polynomial Hierarchy} \label{subsec:hierarchy}
The process of generating the training set outlined in section \ref{subsec:train}
has the added benefit of giving us a set of power spectra constrained around the
peak of the likelihood.  We would like to make use of these points by adding them
to the power spectra training set. However adding a large weighting of points to
a small region of the box will have a negative effet on the accuracy of the 
algorithm outside this region. Instead we have implemented the ability to use a 
hierarchy of polynomials with Pico by separately training over the uniformly
sampled points in the full box and over only the points in the constrained region.
If Pico is given a set of input cosmological parameters within this region it
computes the power spectra based on a polynomial fit to this constrained region.
For points outside this region, Pico defaults to using the polynomials fit over
the full box.
While we have found that using only a single set of polynomials trained on the box
is sufficient for analysis of current experimental data, this
will be a useful feature in the future when data from higher resolution 
experiments become available.

\section{Results} \label{sec:results}
Here we demonstrate the performance of Pico for nonflat cosmologies with 
the dark energy equation of state, $w_{\mathrm{DE}}$ allowed to vary 
(but still constant for a given model).
In this space Pico fits the power spectrum and likelihood as a function of
\begin{equation*}
   \left( \Omega_{\mathrm{b}}h^2, \Omega_{\mathrm{cdm}}h^2, \Omega_{\mathrm{k}},
          \theta, \tau, n_{\mathrm{s}}, \ln 10^{10} A_{\mathrm{s}}, 
          r, w_{\mathrm{DE}}
   \right).
\end{equation*}
The following sections study the accuracy of Pico in computing the power spectra,
matter transfer function, WMAP likelihood as well as its application to parameter
estimation based on this $9$ parameter model.

\subsection{Power Spectra and Matter Transfer Function}
In order to demonstrate Pico's accuracy and robustness we will test the algorithm
for two cases. The first case implements the hierarchy method discussed in
section \ref{subsec:hierarchy}. Here the training set is divided into two pieces.
The first contains $\sim18000$ samples generated uniformly from the box defined in 
Table \ref{tbl:param_bounds}, and the second set consists of the $15000$ points
constrained to $25$ log-likelihoods from the peak of the WMAP likelihood. 
For this case the test set consists of $\sim2000$ points taken from the latter 
training set.  These points were removed from the training set and not
used to train Pico.

The models in the training set were run through CAMB at accuracy settings
($3$,$3$,$3$) to compute the true power spectra and transfer functions.  
As the $\ell$ and $k$ sampling used by CAMB is model dependent it is necessary 
to spline the power spectra and transfer function so that each is computed
at the same $\ell$ or $k$ value. The $\ell$-values were chosen to be those 
used by CAMB for flat models with \texttt{lSamp}$=3$. For the transfer 
function we used a unform sampling in $\ln k$.
Pico was trained using $6^{\mathrm{th}}$ order polynomials,
requiring less than $30$ minutes on a $2.4$GHz desktop.

\begin{figure}
\begin{center}
   \epsscale{1.15} \plotone{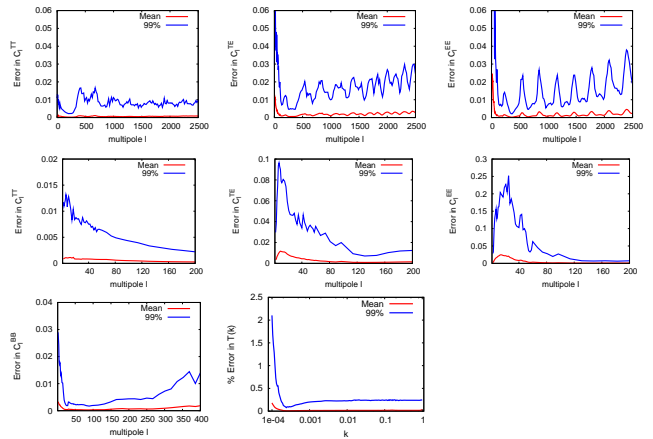}
   \caption{The above plots compare the performance of Pico with CAMB at
            high accuracy settings for 9 parameter nonflat models with 
            $w_{\mathrm{DE}}\ne 1$. Pico was trained using the hierarchy
            method described in section \ref{subsec:hierarchy} and the
            test set consists of $2000$ points within $25$ log-likelihoods
            of the WMAP peak.
            The top two rows show the error 
            compared with CAMB in units of cosmic 
            standard deviation for the TT, TE and EE power spectra at
            high $\ell$ (top) and low $\ell$ (center).
            The bottom row shows the error in the BB spectra in units of 
            the cosmic standard deviation and the percent error in the 
            matter transfer function.
            The two lines on each plot denote the mean error and the error
            bar that bounds $99\%$ of the test set.
            We note that much of the error at low $\ell$ in the EE spectra
            is due to the $1\%$ of models with extremely low power over this range.
            These spectra are too small to detect even with Planck.
            \label{fig:openw}}
\end{center}
\end{figure}

Pico's performance on this test set is shown in Figure \ref{fig:openw}.
The top $2$ rows show the TT, TE and EE power spectra with the second
row focusing on low $\ell$. Results for the BB spectra and matter transfer
function are shown in the third row.
The two lines in each plot represent the mean error and the error bar that bounds
$99\%$ of the test set. For the power spectra the error is plotted in units of
the cosmic standard deviation and for the matter transfer function the $y$-axis 
shows percent error.
From the figure we see that over the volume of parameter space important for
CMB parameter estimation Pico can compute the power spectra for $99\%$ of 
models in the training set to better than $4\%$ of cosmic standard deviation,
with the mean error around $0.5\%$, over most $\ell$-values. 
Even the worst fits to the EE power spectra, which occur just after the 
reionization bump, are only about $25\%$ of the cosmic standard deviation.
We also note that many of the models that are fit poorly have very low power 
in this region so no experiment should be sensitive to these errors.
For the transfer function the $99\%$ error bar
is around $0.25\%$, and the mean at $0.02\%$, except at very low $k$. 
This should be sufficient for analysis of data from the next generation 
of large scale structure experiments.

\begin{table}[ht]
\begin{center}
\begin{tabular}{|ccccc|}
   \hline
   $0.018$ & $<$ & $\Omega_\mathrm{b} h^2$   & $<$ & $0.034$ \\
   $0.06$  & $<$ & $\Omega_\mathrm{cdm} h^2$ & $<$ & $0.2$   \\
   $-0.3$  & $<$ & $\Omega_\mathrm{k}$       & $<$ & $0.3$   \\
   $1.02$  & $<$ & $100\,\theta$             & $<$ & $1.08$  \\
   $0.01$  & $<$ & $\tau$                    & $<$ & $0.55$  \\
   $0.85$  & $<$ & $n_{\mathrm{s}}$          & $<$ & $1.25$  \\
   $2.75$  & $<$ & $\ln \left(10^{10} A_{\mathrm{s}} \right) $ & $<$ & $4.0$ \\
   $0$     & $<$ & $r$                       & $<$ & $2$     \\
   $-1.5$  & $<$ & $w_{\mathrm{DE}}$         & $<$ & $-0.3$  \\
   \hline
\end{tabular}
\end{center}
\caption{Parameter bounds defining the box the training set was sampled 
         from for the example in section \ref{sec:results}. This encompasses a volume
         of at least $3\sigma$ in each parameter around the WMAP maximum likelihood.
         Note that we also impose the prior that the corresponding Hubble constant for each 
         parameter point lie in the interval $\left[30,100\right]$ which excludes some 
         regions inside the box. 
         \label{tbl:param_bounds}}
\end{table}

\begin{figure}
\begin{center}
   \epsscale{1.15} \plotone{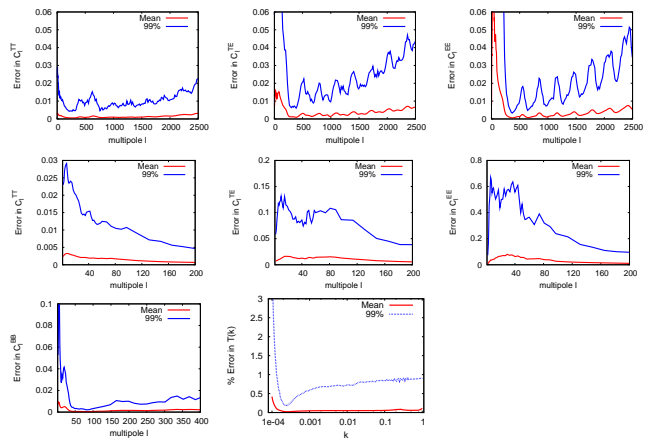}
   \caption{The plots are same as those in Figure \ref{fig:openw}
            except here Pico was trained and tested over models
            sampled uniformly from the box defined by 
            Table \ref{tbl:param_bounds}. Even over this larger
            region Pico can compute the power spectrum in $99\%$
            of the test cases to better
            than $5\%$ of cosmic standard deviation over most $\ell$
            and is never worse than $0.7$ cosmic standard deviation.
            \label{fig:openw-box}}
\end{center}
\end{figure}

For the second test case the hierarchy method is not used and Pico is only
trained on a uniform sample of points from the box in 
Table \ref{tbl:param_bounds}. For this case the test set consists of 
a uniform sample of $\sim2000$ points from the same box. 
Pico's performance on this test set is shown in Figure \ref{fig:openw-box}.
We include this case only to allow comparison with other codes.
When Pico is used to explore the parameter posterior based on CMB constraints, 
which is its main purpose, chains will rarely propose points outside the 
constrained volume used in the hierarchy method.  Therefore 
Figure \ref{fig:openw} provides a better indicator of the types or error
incurred by using Pico to compute the power spectra.
The regression files on the Pico website implement the hierarchy method.

\subsection{WMAP Likelihood}
\begin{figure}
\begin{center}
   \epsscale{1.15} \plotone{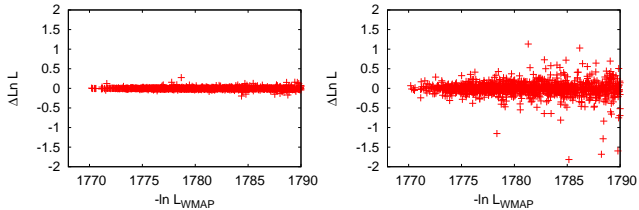}
   \caption{The plots show the absolute error when evaluating the
            WMAP likelihood with Pico.  In the left plot Pico was
            used to compute the power spectrum which were fed into 
            the WMAP code. The right plot shows the absolute error
            when Pico directly evaluates the likelihood. In both 
            cases the likelihood is compared to the value of the 
            WMAP code using high accuracy CAMB power spectra.
            \label{fig:lnlike}}
\end{center}
\end{figure}

Next we test the computation of the WMAP likelihood from Pico for two cases.  The
first uses Pico to compute the power spectrum and then the WMAP code to compute 
the likelihood and the second uses Pico to directly compute the likelihood.
The training set for the likelihood computation consists of $\sim15000$ points generated
using the method described in section \ref{subsec:train}.
Another $\sim2000$ points, generated using the same method, were used as a test set.
The absolute error between the likelihood is shown in Figure \ref{fig:lnlike}. 
The plot on the left shows the results of using Pico to compute the power spectrum
while the plot on the right shows the results of directly computing the likelihood.
For the case of directly evaluating the likelihood, Pico can compute about 
$90\%$ of the test set better than $0.25$ log-likelihoods. When only using Pico to
compute the power spectrum the results are within $0.25$ log-likelihoods for 
better than $99.5\%$ of the models.
The training set and test set were computed using 
high accuracy CAMB runs and version v2p2p2 of the WMAP likelihood 
code.\footnote{http://lambda.gsfc.nasa.gov}

\subsection{Parameter Estimation}
\begin{figure*}[p]
\begin{center}
   \epsscale{1.15} \plotone{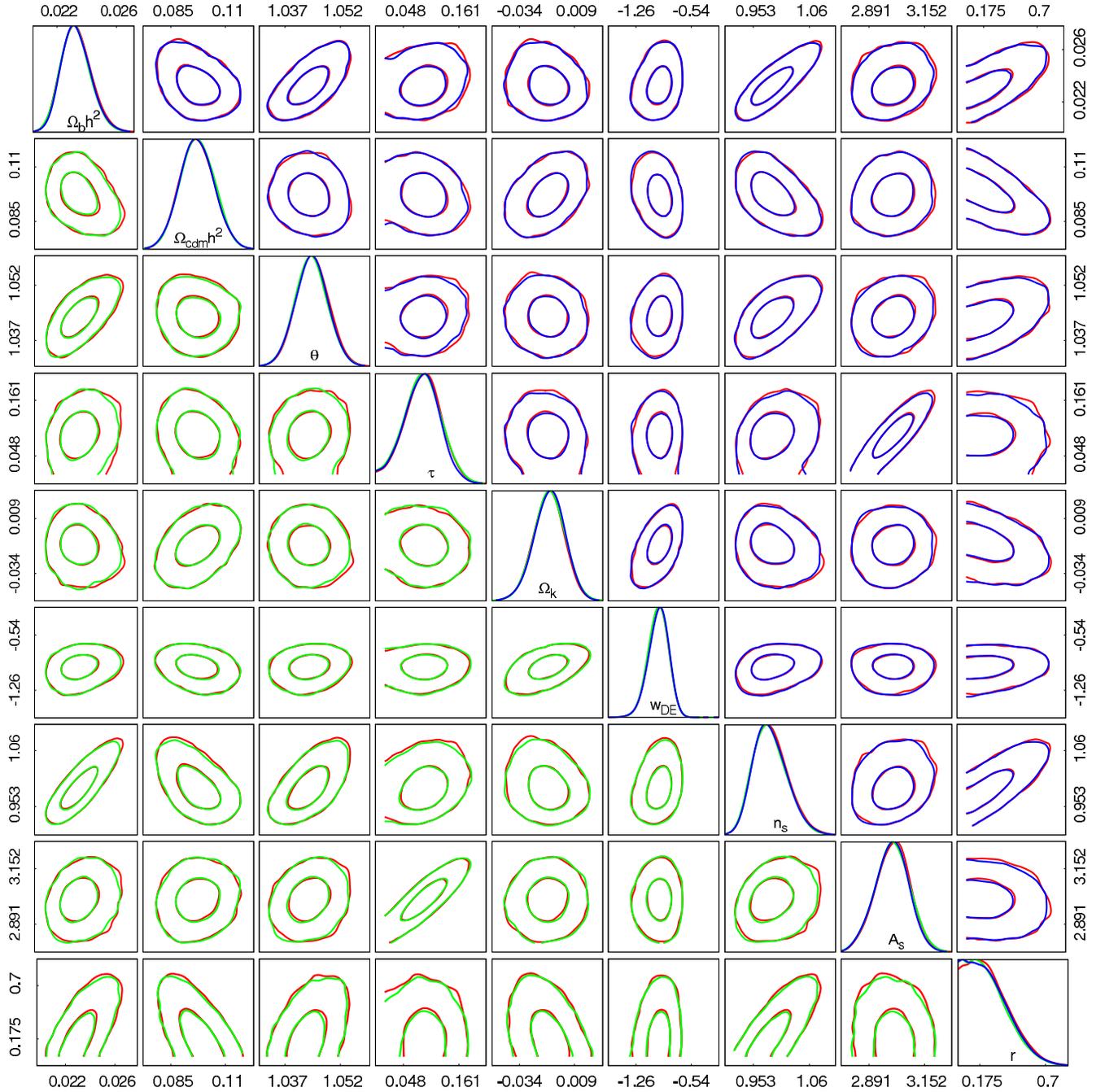}
   \caption{The cosmological parameter posteriors using CAMB and Pico for $9$ parameter
            nonflat models with $w_{\mathrm{DE}}\ne -1$ based on the WMAP, ACBAR, 
            CBI, Boomerang, SDSS, $2$df and SNLS data sets. The red lines are
            the result of using CAMB and the WMAP likelihood code. The green
            lines use Pico to compute the power spectrum but still uses the
            WMAP likelihood. The blue lines are the result from using Pico to  
            compute the power spectrum and WMAP likelihood. The plots in the
            lower triangle show the $68\%$ and $99\%$ contours for the chains
            run using CAMB and PICO with the WMAP likelihood code. The upper 
            triangle compares the CAMB chains to those using Pico to compute 
            the power spectra and WMAP likelihood.
            \label{fig:openw-post}}
\end{center}
\end{figure*}

To test the application of Pico to this $9$ parameter model we ran Markov chains
using CosmoMC with the 
WMAP\citep{Hinshaw:2006ia,Page:2006hz}, 
ACBAR\citep{Kuo:2002ua}, 
CBI\citep{Readhead:2004gy}
Boomerang\citep{Montroy:2005yx,Jones:2005yb,Piacentini:2005yq}, 
$2$df\citep{Cole:2005sx}, 
SDSS \citep{AdelmanMcCarthy:2005se},
and SNLS \citep{Astier:2005qq} 
data sets.
Chains were run for $3$ cases. The first uses CAMB and the WMAP likelihood
code (CAMB$+$WMAP), the second uses Pico to compute the power spectra and
transfer function but still uses the official likelihood codes (PICO$+$WMAP)
and the third case uses Pico to compute the power spectra, transfer function
and the WMAP likelihood (PICO). In third case we did not use Pico to fit
the $2$df, SDSS or SNLS likelihood codes. The $1$-dimensional, marginalized
posteriors for each of the $9$ parameters are shown along the diagonal 
in Figure \ref{fig:openw-post}. The plots in the lower (upper) triangle
in the figure compare the $2$ dimensional posteriors between the 
PICO$+$WMAP (PICO) case and the CAMB+WMAP case. In all of the plots the
CAMB$+$WMAP results are shown in red, the PICO$+$WMAP results in green and 
the PICO results in red.
The lines in the 2D plots denote the $68\%$ and $99\%$ contours.
Using $6$ chains run in parallel, the PICO+WMAP chains ran about $60$ times 
faster than without Pico, requiring about $4$ hours of wall clock time.
Using Pico to compute the WMAP likelihood gave another factor of $2.5$ decrease
in CPU time giving a total speed up of $\sim150$ over chains run without Pico.
In all cases the chains finished with a Gelman-Rubin statistic less than $1.01$.

\section{Conclusion} \label{sec:conclusion}
This paper describes a major new release of Pico, a fast and accurate code for computing
the CMB power spectrum, matter transfer function and the WMAP likelihood. 
We noted the presence of numerical noise
in CAMB at standard accuracy for nonflat models and its effect on the WMAP likelihood.
To solve this problem we have generated training sets running CAMB at high accuracy 
settings.  
Also we have presented a method of generating a training set that finds
the high likelihood region of parameter space without ever running CAMB in serial.  
This is especially useful for training Pico to fit the WMAP likelihood in large
dimensional spaces.
Furthermore, Pico can be trained separately on the power spectra in this smaller region of 
parameter space allowing even more accurate results around the peak of the likelihood
while still maintaining the ability to compute the power spectra over a large box in
parameter space.
The combination of these improvements, along with modifications to the Pico algorithm,
have increased its accuracy in computing the power spectrum and likelihood. 
Also we have extended Pico to compute the power spectrum due to tensor perturbations
as well as the matter power spectrum.
On the Pico homepage,\footnote{http://cosmos.astro.uiuc.edu/pico}
we provide the new version of Pico and new sets of regression coefficients. 
We have also released the training code for Pico, allowing users to apply
the algorithm to new classes of models and parameter sets.
We expect that the accuracy and speed achieved by Pico will be useful for current
and future CMB and large scale structure observations. Furthermore, we hope that
the concept, embodied by Pico, of exploiting massively parallel computing 
resources to solve inherently serial numerical problems will find applications
beyond the immediate domain of cosmological parameter estimation.

\acknowledgements
This work was partially funded by NSF grants AST 05-07676 and AST 07-08849,
by NASA contract JPL1236748, by the National Computational Science Alliance
under AST300029N, by the University of Illinois, by the Computational Science
and Engineering Department at the University of Illinois and by a Friedrich Wilhelm 
Bessel research prize from the Alexander von Humboldt foundation.
We utilized the Teragrid\citep{Catlett}
Itanium 2 clusters at NCSA and at Argonne National Laboratory,
as well as the Turing cluster
in the Computational Science and Engineering Department at the 
University of Illinois at Urbana-Champaign.

We thank the Max Planck Institute for Astrophysics for its hospitality while part
of this work done.
We also thank the users of the Cosmology@Home\footnote{http://www.cosmologyathome.org} 
project whose donated CPU hours 
helped make this work possible.\footnote{http://www.cosmologyathome.org/top\_users.php}
In particular we would like to thank Scott Kruger, the administrator of Cosmology@Home,
as well as the users laurenu2, PoorBoy, 
$\left[\right.$B$\hat{\;\;}$S$\left.\right]$ralfi65, Mitchell, and Mike The Great 
as representatives of all Cosmology@Home participants.  Lastly we would like to thank Nikita Sorokin
for his work in designing the Pico homepage.

Funding for the Sloan Digital Sky Survey (SDSS) has been provided by the Alfred P. Sloan
Foundation, the Participating Institutions, the National Aeronautics and Space Administration,
the National Science Foundation, the U.S. Department of Energy, the Japanese Monbukagakusho, and
the Max Planck Society. The SDSS Web site is \verb+http://www.sdss.org/+.
The SDSS is managed by the Astrophysical Research Consortium (ARC) for the Participating
Institutions. The Participating Institutions are The University of Chicago, Fermilab, the
Institute for Advanced Study, the Japan Participation Group, The Johns Hopkins University, the
Korean Scientist Group, Los Alamos National Laboratory, the Max-Planck-Institute for Astronomy
(MPIA), the Max-Planck-Institute for Astrophysics (MPA), New Mexico State University, University
of Pittsburgh, University of Portsmouth, Princeton University, the United States Naval
Observatory, and the University of Washington.

\pagebreak[4]

\end{document}